\documentclass[a4paper,aps,preprintnumbers,showpacs,amsmath,amssymb,prl]{revtex4}
\usepackage{mathrsfs}
\usepackage{amsmath}
\usepackage[dvips]{graphicx}
\usepackage{epsfig}
\usepackage[dvips]{graphics,graphicx}
\begin{document}

\title{Strongly correlated quantum dynamics of multimode light coupled to a two-level atom in a cavity}

\author{Tarun Kumar$^{1}$, Aranya B.\ Bhattacherjee$^{2}$ and ManMohan$^{1}$}

\address{$^{1}$Department of Physics and Astrophysics, University of Delhi, Delhi-110007, India}\address{$^{2}$Department of Physics, ARSD College, University of Delhi (South Campus), New Delhi-110021, India}

\begin{abstract}
We study a composite multimode light-two-level atom system in a cavity. We show that coupling of the two-level atom to multiple modes of the light destroys the Mott phase of the composite system thus making the system less useful platform for developing concepts in quantum information processing.

\end{abstract}


\maketitle

\section{Introduction}
The search for interesting and potentially useful quantum-mechanical phenomena on a mesoscopic scale in condensed matter and atomic physics is a challenging task. The Jaynes-Cummings model describes the interaction of a single, quasi-resonant optical cavity field with a two-level atom. The optical nonlinearities generated due to the coupling between the atom and the photons leads to an effective photon-photon repulsion \cite{Birnbaum, Imam}. However it was shown that photon-photon repulsion degrades in the presence of many atoms \cite{Grangier,Greentree00}. Later, Rebic et al. \cite{Rebic} showed that the nonlinear interaction afforded by placing a single two-level atom inside a cavity would suffice for realizing photon blockade. Greentree et al. \cite{Greentree} showed that by adding photons to a two-dimensional array of coupled optical cavities each containing a single two-level atom in the photon-blockade regime, a long-lived, strongly interacting dressed states of excitations (coupled atom-photons) are formed which can undergo at zero temperature a characteristics Mott insulator to superfluid quantum phase transition. Devices based on photon blockade mechanism has a strong potential to be useful for quantum computation. In particular because of Mott's phase robustness, devices based on this effect at non-zero temperature has been suggested \cite{Greentree}.
Motivated by such interesting developments in photon blockade schemes, we study in this work, the prospects of utilizing quantum phases of light in a single atom driven by multiple different modes of photons for possible quantum devices. In particular, we consider a two-dimensional array of photonic bandgap cavities. Each cavity contains a single two-level atom, quasi-resonant with the multiple cavity modes. In the classical limit, an array of coupled photonic bandgap cavities has been described for novel waveguide applications \cite{Lang,Ozbay}, nanocavity lasers \cite{Altug} and in the quantum regime a two-cavity arrangement has been proposed as a $Q$-switch \cite{Greentree06}. Recently,the dynamics of linear arrays of coupled cavities containing four-level atomic systems was studied by \cite{Hartmann}.
In the quantum optics context, the generalized Jaynes-Cummings model where the transition is mediated by multiple different modes of photons was first studied by Dantsker \cite{Dan}.

\section{The basic model}

\subsection{Eigenvalues and Eigenfunctions}

To create an atom-multi-photon system whose dynamics we want to study, we consider a two-dimensional array of photonic band gap cavities as discussed earlier \cite{Greentree}. Each cavity contains a single two-level atom, quasi-resonant with the multi-modes of the cavity. To motivate the search for Hubbard-model-type interactions within the present system, we first discuss the eigenvalues, second-order coherence and the probability of finding the atom in the excited state in the presence of multimodes.

The system we consider here is an effective two-level atom with upper and lower states denoted by $|1>$ and $|0>$, respectively. In the multi-photon processes, some intermediate states $|i>$ , $i=2,3..$, are involved, which are assumed to be coupled to $|1>$ and $|0>$ by dipole allowed transitions. Let $\omega_{0}$, $\omega_{1}$, and $\omega_{i}$, denote the corresponding frequency of the atomic energy level $|0>$, $|1>$ and $|i>$, respectively. We also denote $\omega$ as the transition frequency between states $|1>$ and $|0>$. The atom interacts with the $n$ cavity fields with frequencies $\Omega_{i}$, $i=1,2..n$, where $\Omega_{1}+\Omega_{2}+...\Omega_{n}\cong \omega$. Also we assume that detuning between the atomic transition frequency and any one of the $n$ modes is non-zero. Under these circumstances, the intermediate states can be adiabatically eliminated and the effective Hamiltonian of the two-level atom can be written in the rotating-wave approximation as

\begin{equation}
 \hat H^{JC}=\hat H_{0}+\hat H_{I}
\end{equation}

with 

\begin{equation}
 \hat H_{0}=\dfrac{\omega}{2} \hat \sigma_{z}+\sum_{i=1}^{n}\hbar \Omega_{i} \hat a_{i}^{\dagger} \hat a_{i},
\end{equation}

\begin{equation}
 \hat H_{I}=\left( \prod_{i=1}^{n} g_{i}^{1/n} \hat a_{i}^{\dagger} \hat \sigma_{-}+\prod_{i=1}^{n} g_{i}^{1/n} \hat a_{i} \hat \sigma_{+} \right).
\end{equation}

Here $\hat a_{i}^{\dagger}$ and $\hat a_{i}$ are the creation and destruction operator of the $n^{th}$ cavity mode respectively. and $\hat \sigma_{z}$, $\hat \sigma_{-}$ and $\hat \sigma_{+}$ are the usual pauli spin matrices. In the bare state basis $|1,n_{1},n_{2},...n_{n}>$ and $|0,n_{1}+1, n_{2}+1,..n_{n}+1>$, the eigenvalues and the eigenvectors are calculated as

\begin{equation}
 \lambda^{\pm}=K(n_{1},n_{2},..n_{n})\pm Q(n_{1},n_{2},...n_{n}),
\end{equation}

where

\begin{equation}
 K(n_{1},n_{2},..n_{n})=\dfrac{\omega-\Delta}{2}+\sum_{i=1}^{n} \omega_{i} n_{i},
\end{equation}

\begin{equation}
 Q(n_{1},n_{2},..n_{n})=\sqrt{\dfrac{\Delta^2}{4}+\prod_{i=1}^{n} (g_{i})^{2/n}(n_{i}+1)}.
\end{equation}

Here $\Delta=\omega-\sum_{i} \omega_{i}$. The corresponding eigenvectors are

\begin{equation}
|\pm,n_{1},n_{2},..n_{n}>=A^{\pm}_{n_{1},n_{2},..n_{n}}|0,n_{1},n_{2},..n_{n}>+ B^{\pm}_{n_{1},n_{2},..n_{n}}|1,n_{1},n_{2},..n_{n}>,
\end{equation}

where, 

\begin{equation}
 A^{\pm}_{n_{1},n_{2},..n_{n}}=\dfrac{-\Delta /2 \mp Q}{\sqrt{2Q^2 \pm \Delta Q}}
\end{equation}

and

\begin{equation}
 B^{\pm}_{n_{1},n_{2},..n_{n}}=\dfrac{\prod_{i} (g_{i})^{1/n}\sqrt{(n_{i}+1)}}{\sqrt{2Q^2 \pm \Delta Q}}
\end{equation}

\begin{figure}[t]
\hspace{-0.8cm}
\includegraphics[scale=0.8]{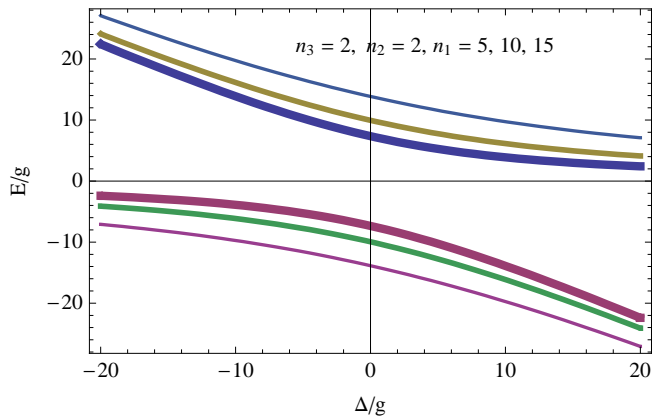}
\caption{Dimensionless eigenspectrum ($E= \lambda^{\pm}-$ optical energy ) for a single atom interacting with 3-modes of a high $Q$-cavity as a function of $\Delta/g$ (dimensionless atom-cavity detuning), centered around zero ($\lambda^{\pm}=\sum_{i=1}^{3}\omega_{i}n_{i}$). The eigenspectrum splits into two branches corresponding to the dressed states, $|+,n_{1},n_{2},..n_{n}>$ (upper branch) and $|-,n_{1},n_{2},..n_{n}>$ (lower branch). The two branches anti-cross at $\Delta/g=0$, with the splitting increasing with increasing excitation number $n_{1}=5,10,15$ (in order of decreasing thickness of the plots)}
\label{1}
\end{figure}

Fig. 1 shows a plot of the dimensionless eigenvalues for the 3 mode case. The photon energies have been subtracted for ease of comparison. The photon number in the first mode ($n_{1}$) is allowed to change while keeping the photon number in the other two modes ($n_{2},n_{3}$) constant. The on-site photonic repulsion is evinced by the increasing energy separation with $n_{1}$. The atom-photon coupling constants $g_{i}$ have been taken to same for all modes.

\subsection{Population probability and Coherence function}

The results of the previous subsection leads to the following matrix representation of the time evolution operator for the given manifold $M(n_{1},n_{2},..n_{n})$.

\begin{equation}
 \hat U(n_{1},n_{2},..n_{n};t)=\left( \begin{array}{cc} \left[ Q+\dfrac{\Delta}{2}\right]e^{-i \lambda^{+}t}+\left[Q-\dfrac{\Delta}{2} \right] e^{i \lambda^{-}t}   &   \prod_{i}(g_{i})^{1/n} \sqrt{n_{i}+1}(e^{-i \lambda^{+}t}-e^{-i \lambda^{-}t})\\   \prod_{i}(g_{i})^{1/n} \sqrt{n_{i}+1}(e^{-i \lambda^{+}t}-e^{-i \lambda^{-}t})   &   \left[ Q-\dfrac{\Delta}{2}\right]e^{-i \lambda^{+}t}+\left[Q+\dfrac{\Delta}{2} \right] e^{i \lambda^{-}t} \end{array} \right) 
\end{equation}

The time evolution operator can now be utilized to calculate the density operator of the system at any time $t$ with an arbitrary initial condition $\hat \rho(0)$ as

\begin{equation}
 \hat \rho(t)=\hat U(t) \hat \rho(0) \hat U^{\dagger}(t)
\end{equation}

The expectation value of any operator can then be obtained easily as

\begin{equation}
 <\hat O(t)>=Tr[\hat \rho(t) \hat O(o)]
\end{equation}

We assume that at $t=0$ the density operator can be decomposed into its atomic and field parts, i.e,

\begin{equation}
 \hat \rho(0)=\hat \rho^{A}(0) \otimes \hat \rho ^{F}(0),
\end{equation}

where the atom is in the $ith$ energy eigenstate $|i>$ and the fields are in a general state as

\begin{equation}
 |f>=\sum_{n_{1},n_{2},..n_{n}}R_{n_{1},n_{2},.._{n}}|n_{1},n_{2},..n_{n}>.
\end{equation}

\begin{figure}[t]
\begin{tabular}{cc}
\includegraphics[scale=0.8]{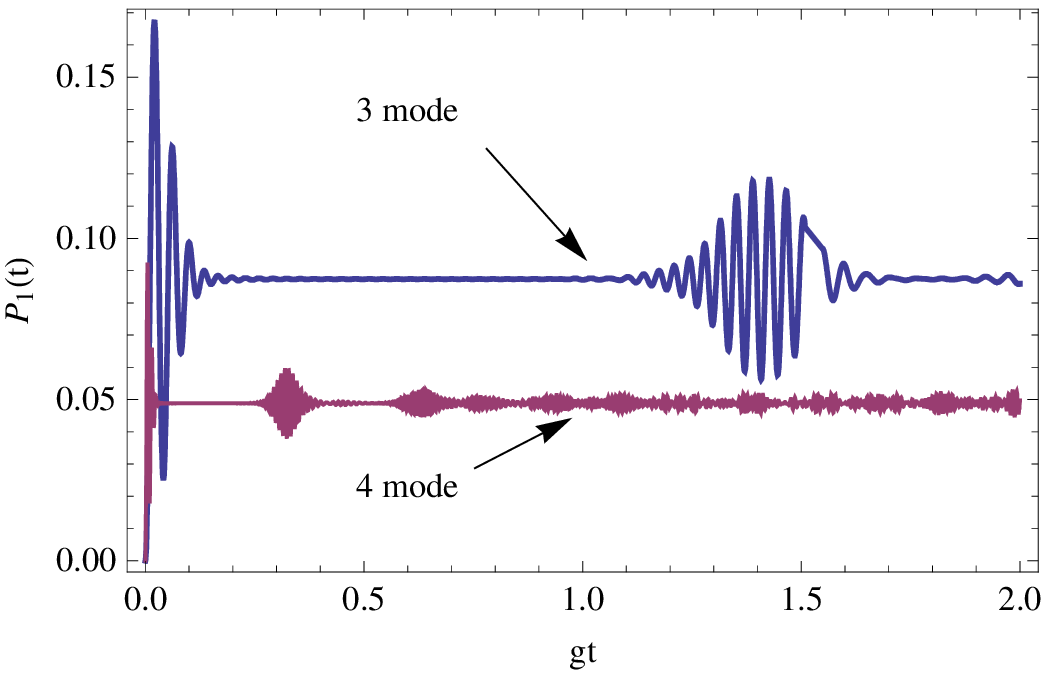}& \includegraphics[scale=0.8]{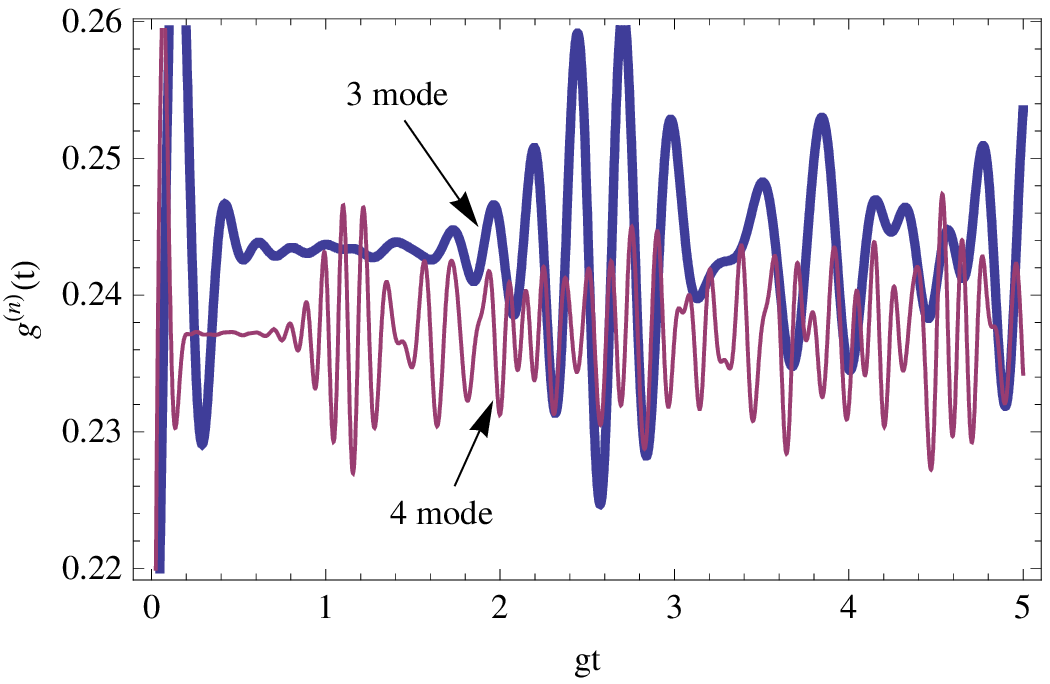}
\end{tabular}

\caption{ Probability $P_{1}(t)$ (left plot) of finding the atom in the excited state and coherence function (right plot) as a function of $gt$. The initial field states are prepared in 3-modes and 4-modes of coherent states with mean photon number $\bar n_{i}=20$ in each mode. }
\label{2}
\end{figure}

Futhermore the reduced atomic density operator $\hat \rho^{A}(t)$ and the reduced field density operator $\hat \rho^{F}(t)$ are introduced by taking the trace of $\hat \rho(t)$ over the field states and over the atomic states respectively. Now if we denote $P_{n_{1},n_{2},..n_{n}}(0)=|R_{n_{1},n_{2},..n_{n}}|^{2}$ as the initial photon distribution, then the probability of finding the atom in the excited state $P_{1}(t)$ for $|i>=|0>$ is calculated as

\begin{equation}
 P_{1}(t)= \rho^{A}(t)=\sum_{n_{1},n_{2},..n_{n}} P_{n_{1},n_{2},..n_{n}}(0)\dfrac{\prod_{i} g_{i}^{2/n}(n_{i}+1)\sin^{2}{Qt}}{Q^2}
\end{equation}

The initial photon distribution is assumed to be coherent which implies that $R_{n_{1},n_{2},..n_{n}}$ is written as

\begin{equation}
 R_{n_{1},n_{2},..n_{n}}=\dfrac{exp\left[ -\dfrac{\sum _{i} \bar n_{i}}{2}\right] \prod_{i} (\bar n_{i})^{n_{i}/2} }{\sqrt{\prod_{i} n_{i}!}}
\end{equation}

Here $\bar n_{i}$ is the mean photon number in the $i^{th}$ mode. The degree of $n^{th}$ order coherence with zero-time delay is defined as 

\begin{equation}
 g^{(n)}=\dfrac{<\prod_{i=1}^{n} \hat n_{i}(t)>}{\prod_{i=1}^{n} <\hat n_{i}(t)>},
\end{equation}

whose magnitude controls the $n$-photon transition rate.

If we assume the atom to be initially in the excited state then mean photon number of the $ith$ mode and the expectation value of the correlation of photon numbers for the $n$ modes that is related to the degree of second order coherence is

\begin{equation}
 <\hat n_{i}(t)>=Tr[\hat \rho(t) \hat n_{i}],
\end{equation}

\begin{equation}
 <\hat n_{1} \hat n_{2}...\hat n_{n}>=Tr[\hat \rho(t) \hat n_{1} \hat n_{2}...\hat n_{n}].
\end{equation}

The time evolution of the probability of finding the atom in the excited state ($P_{1}(t)$) and the $n^{th}$ order coherence function ($g^{n}$) in the presence of 3-modes and 4-modes of the coherent state are plotted in Fig.2. As seen from the figure the quantum revival and collapse of the Rabi oscillations are more compact for the 4-mode case than the 3-mode case. The more interesting observation is the fact that $P_{1}$ is less for the 4-mode case than the 3-mode case. Similarly the coherence function $g^{4}<g^{3}$. This result has some interesting consequence for the quantum phase diagram of the composite photon-atom system to be calculated in the next section.

\section{Quantum phases of light}

The Hamiltonian of our system is given by a combination of the Jaynes-Cummings Hamiltonian with photon hopping between cavities and the chemical potential term. Here we assume that only one mode ($k^{th}$) mode is able to hope between the cavities. The hopping is achieved by evanescent coupling between the cavities. The tunneling frequency is approximately given by $\kappa=\Omega_{k}/Q$($\Omega_{k}$ is the frequency of the $k^{th}$ mode and $Q$ is the quality factor of the cavities ). If $\Omega_{k}$ is kept large as compared to the frequencies of the other mode then one can achieve tunneling of only the $k^{th}$ mode.

\begin{equation}
H=\sum_{l}H^{JC}_{l}-\kappa \sum_{l} \hat a^{\dagger}_{k,l} \hat a_{k,l\pm 1}-\mu \sum_{l}\left( \sum_{i} \hat a^{\dagger}_{i,l} \hat a_{i,l}+\sigma^{+}_{l} \sigma^{-}_{l}\right) 
\end{equation}

Here $l$ is the site index. We are assuming nearest neighbour hopping and the hopping frequency $\kappa$ is assumed to be same. The chemical potential is same for all site if we assume zero disorder. Here $\left( \sum_{i} \hat a^{\dagger}_{i,l} \hat a_{i,l}+\sigma^{+}_{l} \sigma^{-}_{l}\right)$ is the total number of atomic and photonic excitations(the conserved quantity in our system).
We introduce a superfluid order parameter $\psi=<\hat{a}_{l}>$, which we take to be real and use the decoupling approximation,$\hat{a}_{l}^{\dagger}\hat{a}_{m}=<\hat{a}_{l}^{\dagger}>\hat{a}_{m}+<\hat{a}_{l}>\hat{a}_{m}^{\dagger}-<\hat{a}_{l}^{\dagger}> <\hat{a}_{m}>$. The resulting effective mean-field Hamiltonian can be written as a sum over single sites

\begin{equation}
H^{MF}= \sum_{l}H^{JC}_{l}-z \kappa \psi \sum_{l} (\hat a^{\dagger}_{k,l}+\hat a_{k,l})+z \kappa |\psi|^2 -\mu \sum_{l}\left( \sum_{i} \hat a^{\dagger}_{i,l} \hat a_{i,l}+\sigma^{+}_{l} \sigma^{-}_{l}\right)
\end{equation}

Here $z=3$ is the number of nearest neighbours. To obtain the system's zero-temperature phase diagram, we use the procedure of ref. \cite{Oosten}. All energies are now scaled with respect to $z \kappa$.The unperturbed ground state energy of the state with exactly $\left( \sum_{i} \hat a^{\dagger}_{i,l} \hat a_{i,l}+\sigma^{+}_{l} \sigma^{-}_{l}\right) $ particles is $\lambda^{0}=\bar{\lambda}^{-}-\bar{\mu}\left( \sum_{i} \hat a^{\dagger}_{i,l} \hat a_{i,l}+\sigma^{+}_{l} \sigma^{-}_{l}\right) $, where $\bar{\lambda}^{-}=\lambda^{-}/z\kappa$. We only need to consider the negative branch for the purpose of determining the ground state since $\lambda^{-}<\lambda^{+}$. A change in the total number of quasi-excitations per site will occur when $\bar{\lambda}^{-}_{n_{k}+1}-\bar{\mu}(n_{k}+1)=\bar{\lambda}^{-}_{n_{k}}-\bar{\mu}n_{k}$. We can determine the critical chemical potential, $\bar{\mu}_{c}(n_{k})$, where the system will change from $n_{k}$ to $n_{k}+1$ quasi-excitations per site as

\begin{equation}
\bar{\mu}_{c}(n)=\bar \lambda^{-}_{n_{k}+1}-\bar \lambda^{-}_{n_{k}},
\end{equation}

The second order correction to the energy with the well known expression

\begin{equation}
E_{m}^{(2)}=\sum_{n_{i} \neq m_{i}} \dfrac{|<-,m_{1},m_{2},..m_{n}|V|-,n_{1},n_{2},..n_{n}>|^{2}}{\lambda_{m_{k}}^{(0)}-\lambda_{n_{k}}^{(0)}},
\end{equation}

where 

\begin{equation}
 V=-\left( \hat a^{\dagger}_{k}+\hat a_{k} \right) 
\end{equation}

Here we take the unperturbed eigenvector $|-,n_{1},n_{2},..n_{n}>$ corresponding to $\sum_{i}n_{i}$ particles. Since the interaction $V$ couples only to states with one more or less excitations than in the ground state, we find

\begin{equation}
E_{m}^{(2)}=\dfrac{\left( A^{-}_{m_{k}}A^{-}_{m_{k}-1}\sqrt{m_{k}}+B^{-}_{m_{k}}B^{-}_{m_{k}-1}\sqrt{m_{k}-1}\right)^{2} }{\bar \lambda^{-}_{m_{k}}-\bar \lambda^{-}_{m_{k}-1}-\bar \mu}+\dfrac{\left( A^{-}_{m_{k}}A^{-}_{m_{k}+1}\sqrt{m_{k}+1}+B^{-}_{m_{k}}B^{-}_{m_{k}+1}\sqrt{m_{k}}\right)^{2} }{\bar \lambda^{-}_{m_{k}}-\bar \lambda^{-}_{m_{k}+1}+\bar \mu}
\end{equation}

\begin{figure}[t]
\begin{tabular}{cc}
\includegraphics[scale=0.8]{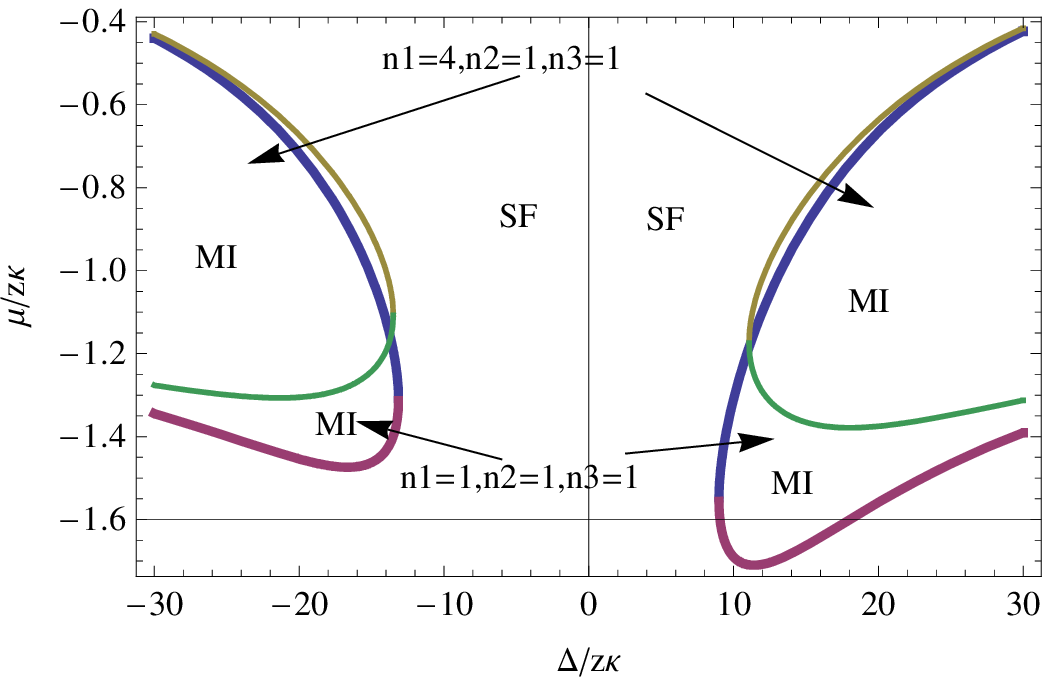}& \includegraphics[scale=0.8]{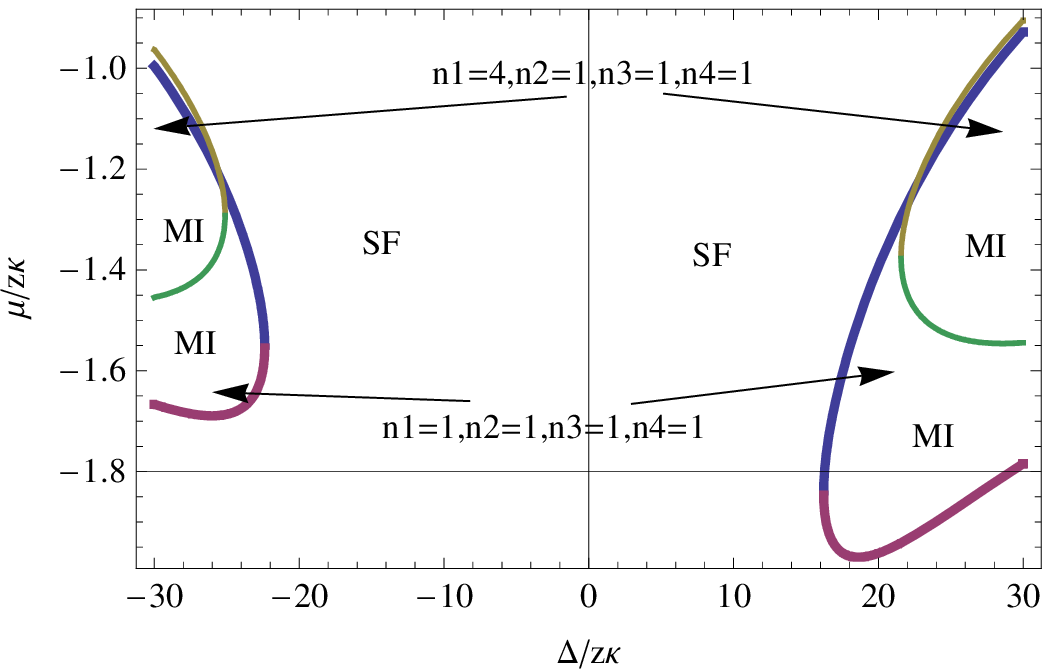}
\end{tabular}

\caption{Phase diagram of the coupled multimode light and the two-level atom syatem in a cavity. The left plot is for 3-mode case and the right plot is for 4-mode case. Dominating the extreme right-hand and the extreme left-hand edge (where photonic repulsion dominates at higher detuning) is the Mott-insulator (denoted by MI) and the superfluid phase is found in between the MI regions (denoted by SF).The size of the Mott region is found to decrease with the quasi-excitations for both the 3-mode and 4-mode case.The size of the Mott lobes is found to be smaller for the 4-mode case as compared to the 3-mode case.}

\label{3}
\end{figure}

According to the Landau procedure for second-order phase transitions, we write the ground state as an expansion in $\psi$

\begin{equation}
E_{m}(\psi)=a_{0}(m,\bar{\mu},\bar{\Delta})+a_{2}(m,\bar{\mu},\bar{\Delta})\psi^{2}+O(\psi^{4}),
\end{equation}

$E_{m}(\psi)$ is minimized as a function of the superfluid order parameter $\psi$. We find that $\psi=0$ when $a_{2}>0$ and that $\psi \neq 0$ when $a_{2}<0$. This means that $a_{2}=0$ signifies the boundary between the superfluid and insulator phases of light. This yields

\begin{equation}
\mu_{\pm}=\dfrac{1}{2}\left( f_{1}^{2}-f_{2}^{2}+\delta \lambda_{1}-\delta \lambda_{2}\right)\pm \dfrac{1}{2} \sqrt{(f_{1}^{2}-f_{2}^{2})^{2}+(\delta \lambda_{1}+\delta \lambda_{2})^{2}+2 (f_{1}^{2}+f_{2}^{2})(\delta \lambda_{1}+\delta \lambda_{2})} 
\end{equation}

where

\begin{equation}
 f_{1}=\left( A^{-}_{m_{k}}A^{-}_{m_{k}-1}\sqrt{m_{k}}+B^{-}_{m_{k}}B^{-}_{m_{k}-1}\sqrt{m_{k}-1}\right)
\end{equation}

\begin{equation}
 f_{2}=\left( A^{-}_{m_{k}}A^{-}_{m_{k}+1}\sqrt{m_{k}+1}+B^{-}_{m_{k}}B^{-}_{m_{k}+1}\sqrt{m_{k}}\right)
\end{equation}

\begin{equation}
 \delta \lambda_{1}=\lambda^{-}_{m_{k}}-\lambda^{-}_{m_{k}-1}
\end{equation}

\begin{equation}
 \delta \lambda_{2}=\lambda^{-}_{m_{k}}-\lambda^{-}_{m_{k}+1}
\end{equation}

The subscript $\pm$ in Eqn. (27) denotes the upper and lower halves of the Mott insulating regions of phase space. Fig. 3 shows the plot of Eqn. (27) for the 3 mode case (left plot) and 4 mode case (right plot). By equating $\bar{\mu}_{-}$ and $\bar{\mu}_{+}$ we can find the point of largest $\Delta_{max}$ for each Mott region (MI). The dynamics illustrated in Fig.3 is extremely rich. Dominating the extreme right-hand and the extreme left-hand edge (where photonic repulsion dominates at higher detuning) is the Mott-insulator (denoted by MI) and the superfluid phase is found in between the MI regions (denoted by SF).The size of the Mott region is found to decrease with the quasi-excitations for both the 3-mode and 4-mode case. The transition from the $SF$ to the $MI$ phase of the quasi-excitations occurs at $\Delta_{max}$. The size of the Mott lobes is found to be smaller for the 4-mode case as compared to the 3-mode case. This observation can be correlated with the observations noted in Fig.2. For the 4-mode case, the probability of transition to the excited state is less as compared to the 3-mode case. The photonic repulsion is a nonlinear process and is directly dependent on the probability of transition to the excited state. A lower transition probability implies a decreased photonic repulsion and hence a smaller Mott region. To find the eigenenergies and to experimentally identify the various phases of the coupled atom-photon system, one can perform a transmission spectroscopy with the scattered light by direct read out of the number of photons coming out of the cavity. Photon loss can be minimized by using high $Q$ cavities and thus ensuring that the light field remains quantum-mechanical for the duration of the experiment.

\section{Conclusions}

In summary, we have studied the strongly correlated dynamics of a two-level atom coupled to multi-mode cavity photons in the photon blockade regime. We have shown that coupling of the two-level atom to multiple modes of the light destroys the Mott phase of the composite system thus making the system less useful platform for developing concepts in quantum information processing.

\section{Acknowledgements}

One of the authors Tarun Kumar acknowledges the Council of University Grants Commission, New Delhi for the financial support under the Junior Research Fellowship scheme Sch/JRF/AA/30/2008-2009.

\end{document}